\documentstyle[psfig,11pt]{article}
\begin{document}
\vspace{1.0cm}
{\Large \bf ASCA observation of the Cygnus Loop Supernova Remnant}

\vspace{1.0cm}

E. Miyata$^{1,2}$, H. Tsunemi$^{1,2}$, K. Koyama$^{2,3}$, and Y. Ishisaki$^4$

\vspace{1.0cm}
$^1${\it Dep. of Earth and Space Science, Graduate School of Science,
 Osaka University \\
	 \ \ 1-1, Machikaneyama-cho, Toyonaka, Osaka 560-0043, Japan }\\
$^2${\it CREST, Japan Science and Technology Corporation (JST)}\\
$^3${\it Cosmic Ray Group, Department of Physics, Kyoto University, \\
Kitashirakawa-Oiwake-Cho, Sakyo, Kyoto 606-8502,  Japan} \\
$^4${\it Dep. of Physics, Tokyo Metropolitan University,
1-1 Minami-Osawa, Hachioji, Tokyo 192-03 Japan}\\

\vspace{0.5cm}

\section*{ABSTRACT}
We present here the results of the mapping observation of the Cygnus
Loop with the Gas Imaging Spectrometer (GIS) onboard the ASCA
observatory. The data covered the entire region of the Cygnus Loop.
Spatial resolution of the GIS is moderate whereas the energy resolving
power is much better than those used in the previous observations. The
ASCA soft-band image shows the well-known shell-like feature whereas
the ASCA hard-band image shows rather center-filled morphology with a
hard X-ray compact source at the blow-up southern region.

\section{INTRODUCTION}
The Cygnus Loop is a proto-type shell-like supernova remnant. Its large
apparent size and the low interstellar absorption feature allow us the
detailed investigation of the spatially-resolved plasma with the ASCA
observatory. Previous ASCA observation reveals that hot plasma
containing rich Si, S, and Fe exists at the center region, suggesting
the ejecta in origin (Miyata et al. 1998a). They estimated that only
1~\% of ejecta is still present at the center region, resulting that a
major part of ejecta is distracted inside the shell region. Therefore, a
mapping of the entire region is important to investigate the
distributions of heavy elements.

A part of the ASCA results was summarized in Miyata (1996).  In this
{\it paper}, we report the preliminary result of the complete mapping
observation of the Cygnus Loop with the ASCA GIS.

\section{OBSERVATION AND DATA ANALYSIS}
We observed the Cygnus Loop from the PV phase (April 1993) to AO-5 (June
1997). Total number of observation is 30.  We re-analyzed all data sets
with the {\it ASCA\_ANL}. We excluded all the data taken at elevation
angle below 5$^\circ$ from the night earth rim and 25$^\circ$ from the
day earth rim, a geomagnetic cutoff rigidity lower than 6 GeV c$^{-1}$,
and the region of the South Atlantic Anomaly. We also applied the
'flare-cut' described in the ASCA News letter No.5. Total exposure time
is 180ksec after the data screening. We made four kinds of images for
each region: a total count image, a non-X-ray background image, a cosmic
X-ray background image, and an exposure map. The image of non-X-ray
background was produced with the H02-sorting method in {\it DISPLAY45}
(the detailed description of this method is in Ishisaki (1996)). The
image of cosmic X-ray background was extracted from the LSS survey
data. We also subtracted the non-X-ray background image from the LSS
data to produce the mean cosmic X-ray background image solely. We used
the day earth image to correct the vignetting effect of the ASCA
telescopes and the grid structure of the GIS since the X-ray spectrum of
the day earth is very soft and is quite similar with that of the Cygnus
Loop.  The data were combined into a single image using {\it DISPLAY45}
and {\it DIS45userlib}.

\section{OVERALL STRUCTURE}

We constructed the X-ray images in the energy band of 0.7-1.5 and 1.5-5
keV as shown in figure~\ref{figure:energy-band}. These images were
corrected both for the exposure and the effective area after subtracting
the background properly. The 0.7-1.5 keV image shows a limb-brightening
structure and is similar to the previously well-known image of the
Cygnus Loop (Ku et al. 1984; Aschenbach 1994).  On the contrary, the
1.5-5 keV image shows a center-filled structure rather than the
well-known shell-like structure. We find two bright regions: a compact
source in the southern region (AX J2049.6+2939) and a north-east (NE)
region.  The X-ray spectrum of AX J2049.6+2939 is much harder than those
of shell regions and can be fitted with a power-law function with a
photon-index of 2.1 (Miyata et al. 1998b). Except AX J2049.6+2939, the
hardest emission can be found at the NE region ($\alpha\simeq 313^\circ,
\delta\simeq 31^\circ$; hereafter we call this region as the northern
hot spot).

Hatsukade \& Tsunemi (1990) performed the scanning observation with
Ginga in the energy band above 1.5~keV and found the center-filled
morphology rather than the shell-like morphology for the Cygnus Loop.
The Ginga intensity profile has a maximum at ($l\simeq 74.9^\circ,
b\simeq 8.6^\circ$), which well coincides with the northern hot spot we
found ($l\simeq 74.9^\circ, b\simeq 8.6^\circ$).  The Ginga intensity
profile also showed a tail structure toward the southern region, which
was probably due to AX J2049.6+2939.

\section{HARDNESS RATIO MAP}

Figure~\ref{image:hardness-ratio} shows the hardness ratio map obtained
with 1.5-5 keV band image dividing by 0.7-1.5 keV band image.  Contour
map overlaid was constructed with 0.7-1.5 keV band image. The northern
hot spot clearly extends toward the north. Comparing the contour map of
the 0.7-1.5 keV image, the northern hot spot is just inside both of the
brightest NE limb and of the northern bright shell region. Miyata et
al. (1998c) investigated the radial profile from the NE limb toward the
center region and found that the kTe distribution showed maximum of
$\simeq$ 1~keV at $\simeq$ 0.4 $R_{\rm s}$, where $R_{\rm s}$ is the
shock radius. This hard spot coincides with the hottest region.

There is a hard X-ray emitting region at the center portion of the Loop.
Miyata et al. (1998a) investigated the center portion in detail and
found hot ($\approx$ 0.8 keV) and metal rich plasma. Such plasma account
for the hard X-ray emitting region we found.

\section{SUMMARY}

We summarize results of our preliminary analysis of the entire Cygnus
Loop.

\begin{itemize}
 \item ASCA soft-band image in 0.7-1.5 keV shows the well-known
       shell-like structure. 
 \item ASCA hard-band image in 1.5-5 keV shows rather center-filled
       morphology and well coincides with the Ginga scanning observation.
 \item AX J2049.6+2939 is the hardest compact source inside the
       Cygnus Loop in the ASCA energy band.
 \item There is a hot spot in the ASCA hard-band image. The hot spot is
       located in the inner region of the bright NE limb.
\end{itemize}

\section{REFERENCES}
\vspace{-5mm}
\begin{itemize}
\setlength{\itemindent}{-8mm}
\setlength{\itemsep}{-1mm}

\item[] 
Aschenbach B. 1994, in New Horizon of X-ray Astronomy,
        ed F. Makino, T. Ohashi (Universal Academy Press, Tokyo) p103

\item[]
Hatsukade I., Tsunemi H. 1990, {\it ApJ}, {\bf 362}, 566

\item[] 
Ishisaki, Y. 1996, Ph.D. thesis of Univ. of Tokyo, ISAS RN 613

\item[]
Ku W.H.-M., Kahn S.M., Pisarski R., Long K.S. 1984, {\it ApJ}, {\bf 278}, 615


\item[] 
Miyata, E. 1996, Ph.D. thesis of Osaka Univ., ISAS RN 591

\item[] 
Miyata, E., Tsunemi, H., Kohmura, T., Suzuki, and S., Kumagai, S.
1998a, {\it PASJ}, {\bf 50}, 257

\item[] 
Miyata, E., {\it et al.} 1998b, {\it PASJ},  {\bf 50}, 475

\item[] 
Miyata, E., {\it et al.}, 1998c, in preparation

\end{itemize}

\clearpage

\begin{figure}[htb]
 \caption{\protect{
 The X-ray image of the Cygnus Loop supernova remnant
 observed with the ASCA GIS in the energy band of (a) $0.7-1.5$ keV
 band and (b) $1.5-5$ keV band. Both images were smoothed with a
 Gaussian function of (a) $\sigma=1^\prime$  and (b) $\sigma=2^\prime$}}
\label{figure:energy-band}
\end{figure}

\begin{figure}[htb]
\caption{Hardness ratio map of the Cygnus Loop (1.5-5 keV band / 0.7-1.5
 keV band). Overlaid contour map was constructed with 0.5-1.5 keV band
 image as shown in figure 1. $1-100$\% of the peak brightness is
 linearly divided into 15 levels.}
\label{image:hardness-ratio}
\end{figure}

\end{document}